# Speaker Diarization using Deep Recurrent Convolutional Neural Networks for Speaker Embeddings

*Paweł Cyrta[1], Tomasz Trzciński[1,2], Wojciech Stokowiec[1,3]*

[1]Tooploox, Poland,
[2]Warsaw University of Technology, Poland,
[3]Polish-Japanese Academy of Information Technology, Poland

[1]`pawel.cyrta@tooploox.com`, [2]`t.trzcinski@ii.pw.edu.pl`, [3]`wojciech.stokowiec@pjwstk.edu.pl`

## Abstract

In this paper we propose a new method of speaker diarization that employs a deep learning architecture to learn speaker embeddings. In contrast to the traditional approaches that build their speaker embeddings using manually hand-crafted spectral features, we propose to train for this purpose a recurrent convolutional neural network applied directly on magnitude spectrograms. To compare our approach with the state of the art, we collect and release for the public an additional dataset of over 6 hours of fully annotated broadcast material. The results of our evaluation on the new dataset and three other benchmark datasets show that our proposed method significantly outperforms the competitors and reduces diarization error rate by a large margin of over 30% with respect to the baseline.

## 1. Introduction

Speaker diarization [1] aims at splitting an audio signal into homogeneous segments according to the speaker identity. More precisely, the goal of speaker diarization is to answer the question of "who spoke when" within a given audio stream. Its application is required in many real-life applications related closely to information retrieval, such as speech-to-text transcription [2] or speaker recognition [3]. Use case scenarios of speaker diarization techniques include the analysis of phone conversations, political debates or lectures and conferences, where determining how many people speak and when they are active is crucial for the proper understanding of information.

Although several advancements in developing high-quality speaker diarization algorithms have been made in the recent years [4, 5, 6, 7, 8], there are still many challenges that have to be addressed, e.g. analysis of the overlapping speech or speaker's voice modulations. A typical approach to solving those challenges is to split speaker diarization problem into three parts: voice activation detection ("speech/no speech"), speaker change detection ("same/different speaker") and speaker identification ("speaker A/B/C"). This partitioning allows to simplify speaker diarization problem, nevertheless it introduces several sub-optimal objectives to be optimized across different modules.

This paper address these shortcomings by learning a set of speaker embeddings that can be used throughout the diarization process. We propose learning the embeddings by training recurrent convolutional neural network for a speaker classification task and prove that this leads to good generalization of the embeddings, even for the speakers not included in the training dataset. Moreover, instead of using manually hand-crafted spectral features such as MFCC [9], PLP [10] or RASTA [11], we apply a recurrent neural network directly on the magnitude spectrograms (SFTF) to learn a set of high-level feature representations also referred to as *speaker embeddings*. Our motivation to use SFTF and not raw audio stream stems from the fact that deep architectures were successfully used to learn Mel-like filters from power spectrum [12], while applying them directly on raw audio data did not lead to classification performance improvement [13, 14, 15]. Inspired by recent advancements in the speaker diarization domain achieved with convolutional neural networks (CNNs) [16] and successful applications of recurrent convolutional neural networks (R-CNNs) to other problems, such as image classification [17] or birds' sound classification [18], we propose employing a recurrent neural network architecture to speaker diarization problem. Our motivation to use this approach is based on the observation that the temporal patterns present in audio streams can be better aggregated and interpreted with recurrent architectures, due to the specific feedback embedded in their design.

To show the superiority of our method over the state of the art, we evaluate it on an exhaustive set of three publicly available datasets with over 720 hours of recording and over 1400 speakers. Additionally, we collect a new dataset of fully annotated audio stream of broadcast news material and release it to the public. As far as we know, this is the only publicly available broadcast dataset. We use this dataset to extend the comparison of our method with the competitive approaches and show that speaker embeddings trained on other datasets generalize well to the new dataset. Our evaluation results clearly show that our approach significantly outperforms the state-of-the-art methods by reducing a classification error rate by a large margin of over 30%.

To summarize, the contributions of this paper are threefold:
- We bridge the gap between the raw audio signal and the output of speaker diarization system using deep neural network architecture.
- To the best of our knowledge, this is the first attempt to use recurrent convolutional neural networks to model low-level speaker embeddings.
- Finally, we introduce a new fully labeled dataset of news, interviews and debates that contain over 6 hours of recording that can empower future research in the domain of speaker diarization.

The remainder of this paper is organized in the following manner: First, we discuss related work and state of the art methods. We then present our method along with the input features and deep neural network architecture. Subsequently, we describe our evaluation setup along with the datasets used for experiments. Finally, we show the results of our evaluation along with the conclusions.

## 2. Related work

Over the recent years speaker diarization systems have been successfully used to analyse human speech in various everyday scenarios - from phone calls [19] through business meetings [20] to broadcast news [7]. In this paper, we mainly focus on the last use case, i.e. the goal of our speaker diarization methods is to automatically annotate broadcast TV and radio streams discussing current news and events.

Historically, the research done in the domain of speaker diarization was highly influenced by the methods proposed for speaker verification. Most notable example is the i-vector framework [7], currently considered as a standard speaker recognition system[1]. An i-vector is an information-rich low-dimensional vector extracted from a feature sequence that represents a speech segment, sometimes known as an *audio voice-print* per analogy to fingerprints. The i-vectors are typically computed on the MFCC features [9] which represent both speaker and channel features. For the proper performance of the speaker diarization system based on i-vectors, it is therefore necessary to add a disambiguation phase which relies on PLDA [21] - a clustering method - and the resulting clustering scores. Since the quality of clustering highly depends on the size of the analysed segments, processing short segments of speech with not enough information leads to significant performance drop of the whole system [22].

To address the above-mentioned shortcomings of the i-vector frameworks, several techniques of so-called anchor modeling were introduced [23]. They create a representation of speech utterance from a set of pre-trained speaker models in order to get a likelihood score of each anchor. Those representations are then grouped into characterization vectors that describe individual speakers. The anchor modeling allows to embed the information about the speaker into the speech segment representation and is often referred to as a *speaker embedding* model. Similar embeddings have recently been obtained through deep neural network training [24]. More precisely, a neural network with three hidden layers was applied to Gaussian Mixture Universal Background Model (GMM-UBM) of MFCC features with the goal of speaker classification. The activations of the so-trained neural network were then used as features of previously unseen speakers. Since then, several extensions of this architecture were proposed, including additional LSTM units [25] and temporal-pooling layers [26, 19]. Once again, the activations of the resulting network were used as speaker embeddings.

In this paper we draw an inspiration from these works and propose a deep neural network architecture to create speaker embeddings for speaker diarization system. Contrary to the previous works, we postulate using recurrent convolutional neural networks as our feature extractor. The motivation to use this architecture stems from the fact that recurrent convolutional neural networks were successfully employed in similar acoustic modeling applications [27] and show state-of-the-art performances on many other related classification tasks, such as document [28], image [17], bird songs [18] and music genre classification[29]. Furthermore, we decided to use CQT-grams instead of raw audio signal, as an input to our recurrent convolutional neural network. We motivate this decision by the inferior performances obtained for deep architectures applied on the raw signals [15, 30].

---
[1] http://voicebiometry.org/

## 3. Method

In this section we first define speaker diarization problem and show its relation to speaker embeddings. We then introduce a set of features used as inputs of our method and follow up with a brief description of the proposed deep neural network architecture.

### 3.1. Problem formulation

The goal of a speaker diarization system is to analyse an audio stream and output a set of labels defining the moments when each individual speaker speaks. This can be cast as a classification task, if all the speakers along with their identities are known beforehand. In real-life applications, however, this is rarely the case. We therefore address the speaker diarization problem using a two-step approach. First, we train a neural network in a supervised manner with a goal of speaker classification. Then, we use the pre-trained neural network to extract time-dependent speaker characteristics, the so-called speaker embeddings. More precisely, we use activations from the last layer of neural network as speaker embeddings. We aggregate the sigmoid outputs by summing all outputs class-wise over the whole audio excerpt to obtain a total amount of activation for each entry and then normalizing the values by dividing them with the maximum value among classes. The analysis of those embeddings in time allows the system to detect speaker change and identify the newly appearing speakers by comparing the extracted and normalized embedding with those previously seen. If the cosine similarity metric between the embeddings is higher than a threshold, fixed at 0.4 after a set of preliminary experiments, the speaker is considered as new. Otherwise, we map its identity to the one corresponding to the nearest embedding.

Since the first stage of our approach involves training a neural network for a speaker classification task, we formulate here the training objective of the network. The input of our network is a weighted spectrogram and a set of ground truth labels defining speaker identity. We train the network to minimize the cross-entropy of the predicted and true distributions:

$$L(y, \hat{y}) = -\sum_{i=1}^{N} \sum_{j=1}^{C} y_i^j \log(\hat{y}_i^j), \quad (1)$$

where $y_i^j$ denotes the ground-truth label of sample $i$ and $\hat{y}_i^j$ refers to the prediction probability estimated by the network; $N$ and $C$ indicate the number of training samples and classes respectively.

### 3.2. Features

As mentioned above, the input of our neural network architecture is a weighted spectrogram. More precisely, we use basic time-frequency representation, i.e. a magnitude spectrogram (SFTF), with some perceptual weighting filter banks applied on the spectrogram to increase the method robustness to noise and pitch variability. Although several previous works reported worse performances when using deep architectures with SFTF instead of Mel-spectral features [31, 29], our preliminary results showed that weighting SFTF with proper perceptual weighting filters may overcome those shortcomings. Below we describe the weighting filters that were separately used to test as input to network in our experiments :

- log-Mel: We transform an input spectrogram by applying Short-Term Fourier Transform and compute Mel filter bank features [9] on the spectral output. To that end, we warp the frequency axis to match the Mel frequency scale. The resulting output is a 96-dimensional binned distribution.

- Gammatone: We apply a Gammatone filter [32] to an input spectrogram to obtain Gammatone-grams. Gammatone filter can be interpreted as an approximative model of the sound filtering performed by human hearing system. The resulting output is again a 96-dimensional representation.

- CQT: We apply a Constant Q Transform [33] an input spectrogram to generate the so-called CQT spectrograms (or CQT-grams). The transform is computed on four octaves with 24 bins per octave and minimum frequency of 80 Hz. The resulting output is also a 96-dimensional vector.

Before applying the filter banks on the spectrograms, we apply several preprocessing steps that aim at normalizing the input signal. First, we transform a stereo audio signal to mono by averaging over two channels. We take fixed-length audio signal segments of 3.072 seconds (96 frames of 512 audio samples) as an input to the system. Each segment is extracted every 250 ms, as this sampling frequency was reported to improve clustering [21, 26]. We then downsample the signal to 16 kHz in order to grasp only the frequency range that is most relevant to speaker identification. Afterwards we apply a Hamming window of 512 samples (32 ms) with hop of 256 samples combined with a pre-emphasis filter and Fast Fourier Transform (FFT). All the parameters of the preprocessing steps were selected based on a set of preliminary experiments.

### 3.3. Network architecture

In this section we present the architecture of the proposed recurrent convolutional neural network (R-CNN) which is trained to model speaker embeddings.

Our network consists of 11 learnable layers: 4 convolutional blocks followed by 2 recurrent layers with 1 fully-connected at the end. Each convolutional block consists of a convolutional layer, a batch-normalization layer [34] and ELU nonlinearity, as well as a max-pooling layer.

We select hyperparameters of the convolutional blocks such as the sizes of convolution kernels and a max-pooling operator, based on a set of preliminary results. For each convolutional block we use the same size of convolutional kernel: $3 \times 3$. As far as max-pooling layers are concerned, following kernel sizes have been used $(2 \times 2)$ in the first block, $(3 \times 3)$ in the second, then $(4 \times 4)$ in the third and in the fourth. For max-pooling layer we use stride equal to kernel width to make those operations non-overlapping. We treat the resulting feature map of size $N \times 1 \times 15$ as a sequence of 15 $N$-dimensional vectors and feed them in recurrent layers with GRU gating mechanism [35].

Our preliminary results suggest that employing exponential linear units (ELUs) introduced in [36] instead of ReLUs slightly improves classification accuracy. Contrary to the claim made in [36], we did not observe any speed-ups in learning.

During training we observed an internal covariate shift, i.e. the change in distribution of each layer's inputs resulting from the updates of parameters of the previous layers. This typically slows down the training process and requires lower learning rates along with a careful parameter initialization [34]. To address these problems, we extended our architecture with an additional batch-normalization layer placed after each convolutional layer.

| Dataset | # speakers | # hours |
|---|---|---|
| AMI [39] | 150 | 100 |
| ICSI [40] | 50 | 72 |
| YouTubeSpeakersCorpus (YT) [41] | 998 | 550 |
| Broadcast News Videos (BNV)[2] | 218 | 6 |
| **Total** | **1416** | **728** |

Table 1: *Overview of the datasets used in the experiments. All datasets contain 728 hours of recordings and 1416 speakers.*

We regularize the network with a dropout rate [37] of 0.1 after each convolutional block, and a stronger rate of 0.3 after the recurrent block. We choose dropout rates based on the preliminary results. To minimize the expression form equation 1 we employ Stochastic Gradient Descent with mini-batches of size 128 with ADAM optimization algorithm [38].

## 4. Experiments

In this section we first introduce datasets used to train and evaluate our method. We then we describe implementation details of our method and the baselines. Finally, we present the evaluation metrics along with the results of our experiments.

### 4.1. Datasets

To evaluate our method and compare it with the state of the art, we use following publicly available datasets: AMI meeting corpus [39] (100 hours, 150 speakers), ISCI meeting corpus [40] (72 hours, 50 speakers), and YouTube (YT) speakers corpus [41] (550 hours, 998 speakers). Those datasets cover a wide range of different types of voices of English speakers and various recording quality standards.

To extend our evaluation method, we collected a set of broadcast material from major news stations: CNN, MSNBC, Fox News, Bloomberg, RT America, BBC. For each of the channels we acquired 1 hour of YouTube clips with 3.2 individual speakers on average per clip and 36.4 speakers per channel. The average speech material for an individual speaker is 2 minutes and 12 seconds. The resulting material is over 6 hours long and contains various types of news programs, interviews and debates. It was then manually labeled with speakers names assigned to every speech segment. Since the dataset also contains pointers to videos, it can also be used for speech recognition, face detection and optical character recognition tasks. We release our Broadcast News Videos dataset (BNV) to the public[2]. One shall note that due to the license restrictions on the video material, the dataset contains only tools and annotations and not raw video files. To recreate a complete dataset, one shall use the download scripts available in the repository.

The overview of the datasets can be seen in Tab. 1. In total, the datasets used for evaluation cover over 1400 hours of recording and over 720 individual speakers. Furthermore, for evaluation purposes all datasets except for the Broadcast News Video dataset are randomly split into training and testing sets with the proportion of 70% to 30%. Broadcast News Video dataset is used only for testing since its goal is to verify the performance of our speaker diarization method on previously unseen material. When evaluating the methods on Broadcast News Video dataset, the training is done on a union of all the other datasets, that is on AMI, ISCI and YouTube speakers corpus.

---

[2] http://github.com/cyrta/broadcast-news-videos-dataset

## 4.2. Implementation

We implement our method that uses recurrent convolutional neural network in Python. For acoustic feature extraction, we employ Yaafe toolkit[3] and we derive magnitude, Mel and CQT spectrogram using this software. To build a deep neural network architecture we use Keras [42], together with Theano [43]. Our model was trained on Nvidia Titan X GPU and the training took 18 hours.

## 4.3. Baseline

The baseline for our comparison is a state-of-the-art LIUM Speaker Diarization system [44]. LIUM is based on a GMM classifier and uses 13 MFCC audio features as an input. The features are calculated using frames of 25 ms with a Hamming window and 10 ms overlap. It also relies on CLR clustering of speech segments into speakers.

Additionally, we compare the performance of our method with both convolutional neural network (CNN) and recurrent convolutional neural network (R-CNN) that takes magnitude spectrogram (SFTF) as an input. The goal of this comparison is to validate our hypothesis that using filter banks improves the results with respect to the systems that rely on magnitude spectrograms.

## 4.4. Evaluation metric

As our performance evaluation metric we use Diarization Error Rate (DER) [1]. This metric takes into account both segmentation and classification errors, as well as errors from from speech activity detection stage. The Diarization Error Rate is computed as:

$$DER = E_{Spk} + E_{FA} + E_{Miss}, \qquad (2)$$

where $E_{Spk}$ is speaker error, defined as the time assigned to incorrect speakers divided by total time, $E_{FA}$ refers to false alarm speech, defined as amount of time incorrectly detected as speech divided by total time and $E_{MISS}$ refers to miss speech, defined as the amount of speech time that has not been detected as speech divided by total time. The definition of DER also includes acceptance margin of 250 ms which compensate for human errors in reference annotation.

## 4.5. Results

The results of the evaluation can be seen in Tab. 2. Our proposed deep learning architecture based on recurrent convolutional neural network and applied to CQT-grams outperforms the other methods across all datasets with a large margin. Its improvement reaches over 30% with respect o the baseline LIUM speaker diarization method with default set of parameters.

The results of the R-CNN combined with other features, namely log-Mel-grams and Gammatone-grams are au pair with the best performing configuration, while significantly outperforming vanilla convolutional neural networks with the same inputs. In fact, all types of input features combined with the R-CNN architecture give on average 12% improvement with respect to the results of the CNN architecture. These results clearly show that the recurrent character of our proposed method allows to better model speaker embeddings and therefore leads to better overall accuracy of the speaker diarization system.

---
[3]http://yaafe.sourceforge.net/

| Method | Features | Datasets | | | |
|---|---|---|---|---|---|
| | | AMI | ISCI | YT | BNV |
| LIUM [44] | MFCC | 25.1 | 21.6 | 26.3 | 28.5 |
| CNN | SFTF | 22.4 | 20.4 | 24.2 | 24.6 |
| | log-Mel | 19.2 | 18.1 | 21.5 | 22.3 |
| | Gammatone | 18.9 | 17.5 | 21.3 | 22.1 |
| | CQT | 16.7 | 15.3 | 20.1 | 21.4 |
| R-CNN | SFTF | 21.7 | 19.3 | 23.7 | 24.1 |
| | log-Mel | 15.7 | 14.2 | 19.1 | 20.1 |
| | Gammatone | 15.6 | 14.1 | 18.6 | 19.7 |
| | CQT | **15.3** | **13.8** | **17.8** | **19.6** |

Table 2: *Diarization Error Rates obtained for all datasets. Our proposed R-CNN architecture with CQT-gram features clearly outperforms the other methods. The performance improvement over the baseline LIUM method reaches over 30%.*

## 5. Conclusions

In this work we proposed a novel method that learns speaker embeddings for speaker diarization problem using recurrent convolutional neural network architecture. To the best of our knowledge, this is the first approach that employs recurrent convolutional neural networks to model low-level speaker embeddings in the context of speaker diarization. We evaluate our method on three publicly available datasets and a new dataset we collected and released to the public. We prove that by using recurrent deep learning architecture our proposed method is able to outperform the state of the art by a large margin of over 30% in terms of Diarization Error Rate across all evaluated datasets.

Our future research plans include verifying the effectiveness of deep neural network architectures in the context of noise reduction and source separation. We also plan to investigate unsupervised deep learning techniques, such as deep clustering [45] and recursive autoencoders, on large datasets to obtain speaker embeddings that are able to discriminate between bigger corpora of speakers.

## 6. Acknowledgment

The authors would like to thank NowThisMedia Inc. for enabling this research.